\documentstyle[prd,preprint,aps,preprint]{revtex}
\newcommand{\beq}    {\begin{equation}}
\newcommand{\eeq}    {\end{equation}}
\newcommand{\beqarr} {\begin{eqnarray}}
\newcommand{\eeqarr} {\end{eqnarray}}
\newcommand{\no}     {\nonumber}

\newcommand{\di}     {\mbox{d}}
\newcommand{\Dtz}    {\Delta t_0}
\newcommand{\Dt}     {\Delta t}
\newcommand{\rd}     {\partial}

\begin{document}
\draft
\preprint{
\begin{tabular}{r}
JHU-TIPAC 95023
\end{tabular}
}
\title{
Modifications of the Hubble Law
\\
in a Scale-Dependent Cosmology}
\author{
C. W. Kim
\raisebox{.6ex}{a}
\footnote{E-mail cwkim@jhuvms.hcf.jhu.edu},
T.H. Lee
\raisebox{.6ex}{a,b}
\footnote{E-mail thlee@jhup.pha.jhu.edu}
and
J. Song
\raisebox{.6ex}{a}
\footnote{E-mail jhsong@rowland.pha.jhu.edu}}
\address{ \raisebox{.6ex}{a}
Department of Physics and Astronomy\\
The Johns Hopkins University,
Baltimore, MD 21218, U.S.A.}
\address{\raisebox{.6ex}{b}
Department of Physics
\\
Soong Sil University,
Seoul 156-743, Korea}
\maketitle

\begin{abstract}
\setlength{\baselineskip}{.5cm}
We study some observational consequences of a
recently proposed scale--dependent
cosmological model for an inhomogeneous Universe.
In this model the Universe  is pictured as being
inside a highly dense and rapidly expanding shell
with the underdense center.
For nearby objects ($z \ll 1$),
the linear Hubble diagram is shown to
remain valid even in this model,
which has been demonstrated both
analytically and numerically.
For large $z$, we present some numerical results of the
redshift--luminosity distance relation and the behavior
of the mass density as a function of the redshift.
It is shown that
the Hubble diagram in this model for a locally $open$ Universe
($\Omega(t_0, r \sim 0)=0.1$)
resembles that of the $flat$ Friedmann cosmology. This implies
that study of the Hubble diagram cannot uniquely determine
the value of $q_0$ or $\Omega_0$ in a model--independent way.
The model also accounts for the fact that
 $\Omega_0$ is an increasing function of the redshift.
\end{abstract}


\section{Introduction}
One of the basic assumptions of the standard Friedmann cosmology
is the Cosmological Principle  which states
 that the Universe is homogeneous and isotropic\cite{Weinberg}
so that
every point of the Universe is equivalent.
While isotropy has  been reasonably well established
from,  among others, the observation of the cosmic microwave
background radiation by COBE\cite{COBE},
homogeneity has been challenged by various observations
of the Large Scale Structures (LSS).
Recent galaxy redshift surveys such as CfA 1\cite{CfA1},
CfA 2\cite{CfA2}, SSRS1\cite{S1}, SSRS2\cite{CfA1} and the
pencil beam surveys\cite{pencil} have provided evidence for
the LSS such as filaments, sheets,
superclusters and voids, up to $200-300 h^{-1}$Mpc.
The current interpretation based on the observation of the LSS
is that the Cosmological Principle may  not be applicable,
at least, to the local Universe,
although it may be applicable to the Universe on a very large scale
with a  characteristic distance $\lambda$.
However,
one of the most remarkable consequences of the above
galaxy surveys is that the scale of the largest structures
in each survey is comparable with the extent of the survey itself.
Recently, it has been suggested that
from pencil beam surveys\cite{pencil}
as well as from new deep redshift survey (ESP survey) \cite{ESP},
$\lambda$ should be much larger than the
survey limits $\sim 500-600 h^{-1}$Mpc,
implying the absence of any tendency towards homogeneity up
to the present observational limits.
For example, Pietronero\cite{fractal} has attempted
to explain these phenomena
by suggesting that the LSS shows fractal properties of the Universe.
Even though this non--analytical distribution of matter means that
the Universe is not homogeneous,
the local isotropy preserves the fundamental assumption
that every point of the Universe is equivalent.

Unfortunately, however, a rigorous mathematical  description
of such a Universe is extremely
difficult and in practice it is almost impossible.
Therefore,  it is desirable to simplify the description of
this $inhomogeneous$ Universe to the extent that its
analytical study becomes possible in order to see,
at least, qualitative features of the matter distribution
and cosmological consequences.
History of cosmological models for an inhomogeneous Universe
dates back to as early as 1930's.
 Lemaitre, Tolman and Dingle \cite{Lemaitre}
attempted to describe the evolution of the
$fluctuation$ in the mass distribution. Later in 1947, Bondi
\cite{Bondi} elaborated the model and discussed
observational consequences.
In their model, which we shall call Tolman-Bondi(TB) model,
the global Universe is that of the standard Friedmann cosmology,
implying homogeneity over the region of order $\lambda$.
Recently, in the frame work of the TB model,
Moffat and Tatarski\cite{Moffat} studied,
in order to describe the local inhomogeneity,
cosmology of a local void in the globally Friedmann Universe and
its effect on the measurement of the Hubble constant
and  the redshift--luminosity distance relation.
Since the Universe modeled in \cite{Moffat} consists of
many expanding voids
and we happen to be located at the center of one of them,
the shell--crossing singularity occurs,
implying that different shells collide and the comoving
coordinate become meaningless.

Recently, based on the fact that there is no observational evidence of
approaching towards the homogeneity within the survey limit,
another cosmological model\cite{Kim} was proposed,
whose global feature is not asymptotic to the Friedmann cosmology.
In \cite{Kim}, the observable Universe is modeled
as being inside an expanding bubble
with the underdense center and
 matter inside the bubble is
isotropically (but inhomogeneously) distributed when viewed
by an observer located at the center.
They proposed that such a Universe may be described by the following
inhomogeneous metric
\begin{equation}
\di \tau^2	=	\di t^2 - R^2(t,r)[\di r^2+r^2 \di \Omega^2],
\end{equation}
where $R(t,r)$ is the scale-factor,
dependent on $r$ as well as on $t$.
Therefore, homogeneity in the Cosmological Principle is explicitly
violated, whereas isotropy remains intact.
Based on the high degree of isotropy of the cosmic microwave
background radiation measured by COBE\cite{COBE},
it was assumed that the observer
is located at the center of the bubble (or near it),
albeit the return of the pre--Copernican notion.
Whether this picture is correct or not
can only be decided when the results of the model are confronted
with the observation.
In \cite{Kim}, its cosmological consequences were qualitatively
discussed.
For example, the Hubble constant, the density parameter
and the age of the Universe all became scale--dependent,
whereas the analysis in \cite{Moffat} was simplified
to avoid the possible position--dependent age.
Moreover, because of the lack of the
light propagation solution on which every observation is based,
no explicit and testable  cosmological results were derived,
which can be compared with the observation. Therefore,
it is interesting to
examine observational consequences of such a model and compare with
those of the TB model.

In this paper, we first present  a general
redshift--luminosity distance relation for a certain class of
inhomogeneous cosmological models. Then we apply the result to
the model discussed in \cite{Kim} as an example.
The plan of this paper is as follows.
In Section II, we present
the redshift--luminosity distance relation
for the case of one $(t,r)$--dependent scale factor in the metric.
In order to proceed further to derive some specific observable
consequences, we  have chosen the model proposed in \cite{Kim} as an
example and briefly summarize the model
in Section III.
In Section IV, we derive
modified results of the redshift--luminosity distance
relation and show that they reduce to the well-known
relations in the standard Friedmann cosmology for small $z$, i.e.,
for nearby objects.
Section V deals with some numerical results of this model which are
applicable for large $z$.
Also discussed in this Section is the
observed increase of density parameter
with the redshift in the framework of this model.
A brief summary and conclusions are given in Section VI.

\section{Redshift in a General  Metric
with one Scale Factor }

With  the exception  of astronomical neutrinos and possible
future gravitational waves, most of
the cosmological measurements are based on the electromagnetic
waves, which travel along the null geodesic, i.e., $\di \tau=0$.
Considering only the radial propagation ($\di \Omega =0$),
we have, from Eq.(1),
\begin{equation}
\di r = - \frac{\di t}{R(t,r)}~,
\end{equation}
where the minus sign is chosen since $r$ decreases as $t$ increases.
With a given $R(t,r)$,
Eq.(2) appears as a simple first-order differential equation,
yielding $r$ as a function of $t$. However, complexity of
solving this differential equation becomes immediately apparent
because we will be dealing with the case in which $R(t,r)$
is not factorized into a separable form of $a(t)f(r)$.
The boundary condition to be imposed is as follows.
Since we  measure a signal at $r=0$,
the boundary condition is $r(t=t_{received})=0$.
More specifically, we treat  the solution of Eq.(2) as  $r = r(t,t_0)$
which is a function of $t$
with the boundary condition $r(t=t_0, t_0)=0$.

In order to define the redshift,
we consider two successive wave crests, both of which leave $r$ and
reach us ($r=0$) at different times.
Suppose that two wave crests were emitted at time $t$ and $t+\Dt$
and received by us at time $t_0$ and $t_0+\Dtz$, respectively.
Then, from Eq.(2) and the definition of $r(t,t_0)$, we have
\beqarr
r & =& \int_t^{t_0} \frac{\di t'}{R(t', r(t',t_0))}
 \\ \no
& =&	\int_{t+\Dt}^{t_0+\Dtz}
		\frac{\di t'}{R(t',r(t',t_0+\Dtz))}
{}~.
\eeqarr
It is to be noted that for each wave crest a proper
boundary condition has to be applied,
which is explicitly expressed in the form of $r(t,t_{received})$.
Since $\Dt$ and $\Dtz$ are extremely small
compared with the cosmological time scale,
it is sufficient to  consider only up
to the first order in $\Dt$ or $\Dtz$.
Then, we have, from Eq.(3),
\beq
\frac{ \Dt}{R(t,r(t,t_0))}
=	\frac{\Dtz}{R(t_0, r(t_0,t_0))}
	\left[ 1+R(t_0,0)\int_t^{t_0} \di t' \frac{\rd}{\rd  t_0}
		\left( \frac{1}{R(t',r(t',t_0))} \right)
	\right]~~,
\eeq
yielding the defining relation of the redshift, $z$, as
\beq
1+z	\equiv	\frac{\Dtz}{\Dt}
	=\frac{R(t_0,0)}{R(t,r(t,t_0))}
	\left[ 1+R(t_0,0) \int_t^{t_0}\di t'
		\frac{\rd}{\rd t_0}\frac{1}{R(t',r(t',t_0))}
			\right]^{-1}~~,
\eeq
where the boundary condition, $r(t_0,t_0)=0$ has been used.
Here and hereafter ${\rd \over \rd t_0}$ explicitly means
$\frac{\rd r}{\rd t_0} \frac{\rd}{\rd r}$.
In the case of the Robertson--Walker metric where
$R(t,r)\equiv S(t)$,
the scale factor of the standard Friedmann cosmology,
it is easy to see that Eq.(5) reduces
to the well-known relation, $(1+z)=S(t_0)/S(t)$.

In the real observation,
the most important definition of distance is
the {\it luminosity distance\/}.
As is well known, if a source at comoving distance $r$ emits
light at time $t$ and a detector at $r=0$ receives the light
at time $t_0$,
the luminosity distance, $d_L$, of a  source in the standard Friedmann
cosmology is
\beq
d_L = r S(t_0) (1+z) = r S(t) (1+z)^2~~,
\eeq
where the second equality is due to the relation $S(t_0)=S(t)(1+z)$
in the standard Friedmann cosmology.

The luminosity distance in a spherically symmetric but
inhomogeneous  Universe
was first examined by Bondi \cite{Bondi} in 1947.
In order to avoid the non-zero pressure,
however, two different  scale factors
were introduced in the metric, as was originally
done by Lemaitre, Tolman and Dingle \cite{Lemaitre},
\beq
\di \tau^2 = \di t^2 - X^2(t,r) \di r^2 - Y^2(t,r) \di \Omega^2~~.
\eeq
One of the special features of the TB model is that the pressure
is always zero, as originally designed
so as to be applicable in the matter dominated era only.
The situation considered in \cite{Bondi} is such that a standard
source is at the $center$ and an observer at $(t,r,\theta,\varphi)$.
As was shown in \cite{Bondi},  the ratio of the absolute luminosity
to the apparent luminosity is simply given by $Y^2(t,r)(1+z)^2$.
Two comments are in order here.
First, we note that the light source is at the center,
implying that the light propagates out spherically
with constant surface energy density at any given time,
which is, in general, not the case in an inhomogeneous
Universe.
Secondly, $(t,r,\theta,\varphi)$ is
a coordinate of the observer, not of the source.
That is, in the standard cosmology notation,
$Y(t,r)$ physically corresponds  to $rS(t_0)$.

Since we consider the situation in which
the position of the observer  is
located at the center, the
light from its source off the center
does not even propagate outward in a
spherically symmetrical manner.
Following the picture of the Universe in \cite{Kim}
where the Universe is inside a bubble with the underdense center
(where the observer is located)
and with the highly dense shell,
the light would feel attraction toward the shell,
implying that the path of the light propagation is,
in general, not a straight line.
Moreover, its energy   is not uniformly distributed
over the non-spherical shell at any given time.
Nevertheless, since the position of the observer
is fixed at the center (i.e., $r=0$) and
he/she receives the light  that
propagates on a straight line,
we shall use the following definition of luminosity distance
\beq
d_L \equiv r(t,t_0)R(t_0, r(t=t_0,t_0))(1+z)~~,
\eeq
where the $(1+z)$ factor comes  from the correction factor,
$(1+z)^2$, that appears in the relationship
between the absolute luminosity and
the apparent luminosity (hence only
one factor out of $(1+z)^2$ in the distance).
One factor of $(1+z)$ in the luminosity relation is due to
the decrease of  energy
because of the redshift,
the other factor coming from the increase of the time
interval from $\Dt$ to $\Dtz$, which is also just $(1+z)$
by definition.
Of course, the above  definition has to be justified
by performing the coordinate transformation
from the observer to the source in the inhomogeneous Universe.
This, however, is beyond the scope of this paper and thus,
based on its plausibility,
we shall assume its validity in this paper.
We caution the reader that the luminosity distance
should not be simply written as $r(t,t_0)R(t,r(t,t_0))(1+z)^2$,
for $R(t,r(t,t_0))$ is not simply given by $R(t_0,0)$
times a factor, $(1+z)$, as can  be  seen in Eq.(5).

\section{Scale--dependent cosmology}

In order to obtain some specific results of cosmological
consequences of the proposed inhomogeneous metric, Eq.(1),
we shall consider, as an example, the model of \cite{Kim}.
In this Section, we shall briefly summarize the model.
First, given the metric in Eq.(1), in order to accommodate the
$r$-dependence on the Ricci tensors,
the Einstein equation was  also generalized as
\begin{equation}
R^{\mu \nu}-\frac{1}{2}g^{\mu \nu}R=-8\pi [G T^{\mu \nu}](t,r)~,
\end{equation}
where the  $(t,r)$ dependence of the combination, $[GT^{\mu \nu}]$,
was explicitly noted.
When the non-vanishing elements of Ricci tensor calculated
from Eq.(1) are substituted into the
generalized Einstein equation in Eq.(9),
we obtain the following non-vanishing components.
\begin{eqnarray}
3 \frac{\dot{R}^2(t,r)}{R^2(t,r)}-2\frac{R''(t,r)}{R^3(t,r)}
+\frac{R'^2(t,r)}{R^4(t,r)}-4\frac{R'(t,r)}{rR^3(t,r)}
&=&8 \pi [G\rho] \\
2 \frac{\ddot{R}(t,r)}{R(t,r)}+\frac{\dot{R}^2(t,r)}{R^2(t,r)}
-\frac{R'^2(t,r)}{R^4(t,r)}-2\frac{R'(t,r)}{rR^3(t,r)}
&=& -8 \pi [Gp_r] \\
2 \frac{\ddot{R}(t,r)}{R(t,r)}+\frac{\dot{R}^2(t,r)}{R^2(t,r)}
-\frac{R''(t,r)}{R^3(t,r)}
+\frac{R'^2(t,r)}{R^4(t,r)}-\frac{R'(t,r)}{rR^3(t,r)}
&=&- 8 \pi [Gp_{\theta}] \\
2 \frac{\ddot{R}(t,r)}{R(t,r)}+\frac{\dot{R}^2(t,r)}{R^2(t,r)}
-\frac{R''(t,r)}{R^3(t,r)}
+\frac{R'^2(t,r)}{R^4(t,r)}-\frac{R'(t,r)}{rR^3(t,r)}
&=&- 8 \pi [Gp_{\varphi}]
\end{eqnarray}
where dots and primes denote, respectively, derivatives with
respect to $t$ and $r$.
Another non-vanishing Ricci tensor $R_{01}$ yields
\begin{equation}
R_{01} = 2 \left( \frac{\dot{R}'(t,r)}{R(t,r)}
	-\frac{\dot{R}(t,r)R'(t,r)}{R^2(t,r)} \right)
	= - 8\pi [GT_{01}]~~.
\end{equation}
As was discussed in \cite{Kim},
in order to maintain an inhomogeneous matter distribution,
it is essential to keep pressures and $T_{01}$ to be finite
so that $R(t,r)$ is kept from being factored out as $a(t)f(r)$,
in which case the Robertson--Walker metric is recovered.
This feature distinguishes this model from the TB model,
in which pressure was set to be zero to begin with.
Moreover, to avoid the sheer force, it was assumed in \cite{Kim}
that $p_r=p_{\theta}(=p_{\varphi})$. Then
a constraint on $R(t,r)$ is uniquely determined
from Eqs.(11) and (12) as
\beq
R(t,r) = \frac{ a(t) }{1-B(t)r^2}~~,
\eeq
where $a(t)$ and $B(t)$ are positive, arbitrary functions
of $t$ alone.
The negative sign on the right hand side
of Eq.(15) is chosen to avoid a locally closed Universe
(see below Eq.(16)).
Inserting Eq.(15) into Eq.(10) gives
\begin{equation}
\left[  \frac{\dot{R}(t,r)}{R(t,r)} \right]^2
=	\frac{8\pi }{3}[G\rho](t,r) + \frac{4B(t)}
{a^2(t)} ~~.
\end{equation}
The term, $4B(t)/a^2(t)$, was interpreted
as a time-varying vacuum energy density in \cite{Kim}.
It should be noted here that the appearance of this
term, admittedly very surprising,  is a consequence of
the metric in Eq.(1).
Let us briefly discuss  physical implications of Eq.(16).
In our neighborhood (i.e., $r \ll 1$),
the left-hand side of Eq.(16) is reduced to $(\dot{a}/a)^2$,
implying that $a(t)$ represents more or less the scale factor
for our local Universe.
Moreover, since the standard Friedmann cosmology has been
successful in describing our local neighborhood,
any modifications of
the standard cosmology in this model
must be small in the local Universe.

The next question is whether our local  neighborhood is flat or open.
If it is $flat$,
$B(t)$ should be treated, as can be seen on the right--hand side
of Eq.(16), as being small in accord with the assumption
of small modifications on the local neighborhood.
If the local neighborhood is $open$, however,
$B(t)$ itself is not small.
Upon writing $B(t)$ as $[1+b(t)]/4$,
Eq.(16) becomes
\begin{equation}
\left[  \frac{\dot{R}(t,r)}{R(t,r)} \right]^2
=	\frac{8\pi }{3}[G\rho](t,r) + \frac{1}
{a^2(t)} + \frac{ b(t)}{a^2(t)}~~,
\end{equation}
implying that in the local neighborhood
(i.e., for $r \ll 1$, or equivalently
for $(\dot{R}/R)^2 \simeq (\dot{a}/a)^2$),
small is $b(t)$, but not $B(t)$.

It is interesting to mention here that the metric given by
Eq.(1) with the Einstein equation dictates the behavior of
the energy density,
pressure and momentum density of the  Universe.
We first note
that the constraint on $R(t,r)$, Eq.(15), severely restricts
the behavior of $G\rho$, $Gp$ and $GT_{01}$
as functions of $t$ and $r$.
Substituting Eq.(15) into Eqs.(10), (11) and (14),
we obtain the following explicit expressions :
\begin{eqnarray}
\frac{8\pi}{3}G\rho &=&
	\left(\frac{\dot{a}}{a} \right)^2
	-\frac{1}{a^2}-\frac{ b}{a^2}
	+ 2\left(\frac{\dot{a}}{a}\right)\left( \frac{\dot{b}r^2/4}
	{1- \left[ {1+b \over 4} \right]r^2}\right)
		+\left(\frac{\dot{b}r^2/4}
	{1- \left[ {1+b \over 4} \right]r^2} \right)^2
			 \\
8\pi Gp&=&
	\frac{1}{a^2}-
	2\frac{\ddot{a}}{a} - \left( \frac{\dot{a}}{a}\right)^2
	+\frac{b}{a^2}
	- \left( 6 \frac{\dot{a}}{a}+
	2\frac{\ddot{b}}{\dot{b}} \right)
\left(	\frac{\dot{b}r^2/4}
	{1- \left[ {1+b \over 4} \right]r^2}\right)
	 - 5 \left(
	\frac{\dot{b}r^2/4}
	{1- \left[ {1+b \over 4} \right]r^2} \right)^2
			\\
8\pi GT^{01} &=& \frac{\dot{b}r}{a^2} ~~.
\end{eqnarray}

The following comments are in order here.
{}From Eqs.(18) and (19),
it is easy to see that as $ r$ approaches $\sqrt{\frac{4}{1+b}}$,
we have $p/\rho = -\frac{5}{3}$,
implying a bizarre equation of state.
In the standard inflationary scenario
in which a constant vacuum energy
is responsible for generating an inflationary period,
one has $ p/\rho = -1$.
Therefore, in the model under consideration,
an unusual scalar sector with
time-dependent vacuum energy,
which is as yet to be understood,
may be responsible for the inflation
which is much more rapid than the usual inflation.
By using the arbitrariness of $b(t)$, however,
this singularity with the unusual $p/\rho$ ratio
can be pushed far away from the particle
horizon, so that the local value of $p/\rho$
at the present matter dominated era can be made
to be positively small,  as generally expected.
For this reason, in spite of this bizarre behavior,
we shall proceed to use this model as an example
for our following discussions.

\section{Perturbative Approach}

The linear relationship between the redshift and luminosity distance
with a constant coefficient (the Hubble constant)
for nearby objects (for $z \ll 1$) has been well established by
various observations.
In this Section,
we investigate whether or not the redshift--luminosity distance relation
for small $z$ in an inhomogeneous cosmological model discussed
in the previous Section still remains the same as in the standard
cosmology.

\subsection{Locally flat Universe}

We start with the light propagation equation
\beq
\frac{\di r}{\di t}=-\frac{1}{a(t)}+\frac{B(t)}{a(t)}\,r^2,
\eeq
which is determined by  two arbitrary functions of $t$,
$a(t)$ and $B(t)$.
But unfortunately Eq.(21) cannot be solved
analytically because it is non-linear in $r$.
For a locally $flat$ Universe with  small perturbations
to the standard cosmology,
$B(t)$ should be treated as being very small.
Therefore, we have, from Eq.(21),
\beq
r(t,t_0)=
\int_t^{t_0}\frac{\di t'}{a(t')}
	-\int_t^{t_0} \di t' \frac{B(t')}{a(t')}
	\left[
\int_{t'}^{t_0}\frac{ \di t''}{a(t'')} \right]^2
+{\cal O\/}(B^2)~~,
\eeq
yielding
\beq
\frac{\rd}{\rd t_0}\left[ \frac{1}{R(t,r(t,r_0))} \right]=
	-\frac{2 B(t)}{a(t)a(t_0)} \int_t^{t_0}\frac{\di t'}{a(t')}
		+ {\cal O\/}(B^2)~~,
\eeq
where we have used Eq.(15).
Now, the redshift is given by
\beqarr
1+z &\equiv& \frac{\Dtz}{\Dt}
\\ \no
&=& \frac{a(t_0)}{a(t)} \left[	1-B(t)\left(
\int_t^{t_0}\frac{\di t'}{a(t')} \right)^2
+2\int_t^{t_0}\di t' \frac{B(t')}{a(t')}
\int_{t'}^{t_0}\frac{ \di t''}{a(t'')}
	+ {\cal O\/}(B^2) \right]~~.
\eeqarr
In a special case where $B(t)$ is a constant,
which  corresponds to the standard Friedmann cosmology,
the redshift simply reduces to the standard relation
\beq
1+z	=	\frac{a(t_0)}{a(t)}~~,
\eeq
where we have used the relation
\beq
2 \int_t^{t_0}\frac{\di t'}{a(t')}
\int_{t'}^{t_0}\frac{ \di t''}{a(t'')} = \left[
\int_t^{t_0}\frac{\di t'}{a(t')} \right]^2 ~~ .
\eeq
That is, the well-known redshift expression
in the standard cosmology is reproduced, as expected.
Therefore,  this result strongly suggests that $a(t)$ plays a role of,
more or less, the scale factor of the standard Friedmann cosmology.
In the case of a locally flat Universe, therefore,
the behavior of $a(t)$ cannot be much different from that of the
standard Friedmann cosmology with $k=0$,
which is proportional to $t^{2 \over 3}$ in the matter dominated era.
For  mathematical simplicity and illustrative purposes,
we assume that $a(t)$ behaves the same as that in the standard cosmology
and $B(t)$ can be expressed by a simple power law.
That is, we assume that $a(t)$ and $B(t)$ are, in the matter
dominated era, of the form
\beq
a(t)=\alpha t_0 \,
\left( \frac{t}{t_0} \right)
^{ {2 \over 3} }
{}~,~~
B(t)=\beta
\left( \frac{t}{t_0} \right)^n
{}~( n \ge 0)~~,
\eeq
where $\alpha$ and $\beta$ are dimensionless parameters
to be determined and $n$ is set to be non-negative because of
the observed increase of the matter density
as a function of $r$.\cite{Kim}
Substituting Eq.(27) into Eq.(22) yields
\beqarr
r(t,t_0) &=& \frac{3}{\alpha} \left[ 1-
\left( \frac{t}{t_0} \right)^ {1 \over 3} \right]
-\beta\frac{9}{\alpha^3}
\left[ \frac{2}{9 ( n+ {1 \over 3})(n+{2 \over 3})(n+1)}\right.
\\ \no
& &\left.- \frac{1}{(n+{1 \over 3})}
\left( \frac{t}{t_0} \right)^{ n+{1\over 3}}
+\frac{2} { (n+ {2 \over 3})  }
\left( \frac{t}{t_0} \right)^{n+{2\over 3}}
-\frac{1}{(n+1)}
\left( \frac{t}{t_0} \right)^{n+1} \right]
+{\cal O\/}(\beta^2)~~.
\eeqarr
In practice, what is measured is the redshift.
Therefore,  we must express
$( t/t_0)$ in terms of the red shift.
By substituting Eqs.(27) into Eq.(24), we have
\beqarr
1+z &=&
\left( \frac{t}{t_0} \right)^{- {2 \over 3}}\left[
1-\beta \frac{9}{\alpha^2}\left( \frac{t}{t_0} \right)^n
\left\{1- \left( \frac{t}{t_0} \right)^{1\over 3} \right\}^2
\right.
\\ \no
& & \left. + \beta
\frac{2}{ \alpha^2(n+ {1 \over 3})(n+ {2 \over 3}) } \left\{ 1+3
(n+{1 \over 3})
\left( \frac{t}{t_0} \right)^{n+{2 \over 3}}-3 (n+ {2 \over 3})
\left( \frac{t}{t_0} \right)^{n+ {1 \over 3}}\right\} \right]
+ {\cal O\/}(\beta^2)~~,
\eeqarr
yielding
\beqarr
\left( \frac{t}{t_0} \right)
&\equiv& T_{(0)}+\beta T_{(1)}+{\cal O\/}(\beta^2)~~,
\eeqarr
where, for notational simplicity, we define $T_{(0)}(z)$
and $T_{(1)}(z)$ as
\beqarr
T_{(0)}&=&	(1+z)^{- {3 \over 2}}
\\ \no
T_{(1)} &=&{3 \over 2\alpha^2}
\left[ \frac{ 2 }{(n+ {1 \over 3})(n+ {2 \over 3})}
(1+z)^{- {3 \over 2}}+
\frac{18n}{(n+ {1 \over 3})}(1+z)^{- {3 \over 2}(n+{4\over 3})}
\right.
\\ \no
& & \left.
-\frac{9n}{(n+ {2 \over 3})}(1+z)^{- {3 \over 2}(n+{5 \over 3})}
-9(1+z)^{-{3 \over 2}(n+1)}\right]~~.
\eeqarr
Inserting Eqs.(30) and (31) into the right--hand side of Eq.(28) gives
\beqarr
r(t,t_0)
& = &
r_{(0)}+\beta r_{(1)}
+{\cal O\/}(\beta^2)~~,
\eeqarr
where $r_{(0)}$ and $r_{(1)}$ are defined as
\beqarr
r_{(0)}	&\equiv&
{3 \over \alpha}  [1- T_{(0)}^{1 \over 3}]
\\ \no
r_{(1)} &\equiv&
-{1 \over \alpha^3}
	\left[ T_{(0)}^{-{2 \over 3}} T_{(1)} \alpha^2
		 + \frac{2}{ (n+ {1\over3})(n+{2 \over 3})(n+1)}
\right.
\\ \no
& &\left.
-\frac{9}{(n+ {1 \over 3})} T_{(0)}^{ n+{1 \over3}}
+\frac{18}{(n+{2 \over 3})} T_{(0)}^{n+{2 \over 3}}
-\frac{9}{(n+1)} T_{(0)}^{n+1} \right].
~~.
\eeqarr
It is easy to see that the $r_{(0)}$
term reproduces the result
in the standard Friedmann cosmology whereas
$r_{(1)}$ represents a correction term.
It is interesting to note that
for small $z$,
$r_{(1)}$ is  zero up to the second order in $z$,
as can easily be seen by substituting Eq.(31) into Eq.(33).
That is, there is no modification of small $r(t)$ (i.e., for $z\ll 1$)
due to small $B$,
up to the second order.
Since the luminosity distance is $r(t,t_0)R(t_0,0)(1+z)$,
the redshift--luminosity distance relation for small $z$
remains intact, at least, up to the first order
in $B(t)$. This is a consequence of  the plausible assumption that
$a(t)$ has very similar behavior of the scale factor
of the standard Friedmann cosmology in the matter dominated era.

\subsection{Locally Open Universe.}
Before we proceed, we make a brief comment on
the status of the density parameter $\Omega$.
One of the most challenging tasks in the observational
astrophysics is the measurement of the mass density of the Universe,
which is supposed to be a constant at any given time
in the standard Friedmann cosmology.
Various observations, however, indicate that
the mass density indeed appears to increase as we probe farther out
\cite{Schramm}\cite{Kolb}.
{}From direct observations,
the fraction of critical density associated
with  luminous galaxies is $\Omega_{LUM}\leq 0.01$.
When extending the observation to distances beyond
the luminous part of galaxies,
we found that there exist galactic halos which have a mass
corresponding to $\Omega_{HALO} \simeq 0.1$.
On a larger scale such as the Virgo cluster,
modeling the local distortion of the Hubble flow
around the cluster yields $\Omega_{CLUSTER}=$0.1 to 0.2.
Recently, using the redshift measurements for the catalogue of galaxies
by the Infrared Astronomy Satellite (IRAS)\cite{IRAS},
it became apparent that galaxies out to about 100 Mpc
flow towards the Great Attractor with high peculiar velocity.
It was concluded that the observed dynamics on this scale requires
$\Omega_{IRAS} \sim 1 \pm 0.6$.
Based on the above observations,
we have a picture of the Universe in which
our local neighborhood is underdense and
the mass density increases with scale.

Therefore, we present, in this Subsection,
the redshift--luminosity distance
relation for the locally $open$ Universe.
As was discussed before,
$a(t)$ is more or less the scale-factor
of the locally open Universe in the standard Friedmann cosmology.
Therefore, it is more transparent to rewrite $B(t)$
as $[1+b(t)]/4$.
Hereafter, a small perturbation to the locally $open$ Universe
is represented by $b(t)$ rather than by $B(t)$.
It is convenient to
introduce a new coordinate $\Phi$ as defined by
 $r(t,t_0) = 2 \tanh \Phi(t,t_0)$.
The light propagation equation is then reduced to
\beq
\frac{\di \Phi}{\di t}= -\frac{1}{2a(t)}
		+\frac{b(t)}{2 a(t)} \sinh^2 \Phi~~.
\eeq
We consider  two successive wave crests
that leave a comoving coordinate $\Phi$ at time $t$ and $t+\Dt$
and arrive at $\Phi=0$ at times $t_0$ and $t_0+\Dtz$, respectively,
which yields the following equality :
\beqarr
\Phi &=&\int_{t}^{t_0}\di t'
\left[ \frac{1}{2a(t')}-\frac{b(t')}{2a(t')}
\sinh^2 \Phi (t',t_0) \right]
\\ \no
&=&
\int_{t+\Dt}^{t_0+\Dtz}\di t'
\left[ \frac{1}{2a(t')}-\frac{b(t')}{2a(t')}
\sinh^2\Phi(t',t_0+\Dtz) \right]~~,
\eeqarr
from which the redshift relation is given by
\beq
1+z \equiv \frac{\Dtz}{\Dt}
= \frac{a(t_0)}{a(t)} \left[ \frac
{1-b(t)\sinh ^2 \Phi(t,t_0)}
{1-a(t_0) \int_{t}^{t_0}\di t' \frac{b(t')}{a(t')}
	\frac{ \rd \Phi(t',t_0)}{\rd t_0}
	\sinh 2\Phi (t',t_0) } \right].
\eeq
For an open Friedmann Universe where $b(t)=0$,
$(1+z)$ simply becomes  $a(t_0)/a(t)$, which is
 the standard result.
Since the light propagation equation is non-linear,
we will again use the perturbation method
by treating $b(t)$ as being small.
Then, the redshift in this picture becomes
\beq
1+z \simeq
\frac{a(t_0)}{a(t)}\left[
1-b(t)\sinh^2 \int_{t}^{t_0}\frac{\di t'}{2a(t')}
		+ \int_{t}^{t_0}\di t'
		\frac{b(t')}{2a(t')}\sinh \int_{t'}^{t_0}
			\frac{\di t''}{a(t'')}
		+{\cal O\/}(b^2)\right]~~,
\eeq
where we have used the relation,
$\rd \Phi(t,t_0)/\rd t_0 = 1/2 a(t_0)$.
Again to proceed further we need
specific functional forms of $a(t)$ and $b(t)$.
Since $a(t)$ cannot be too
different from the scale factor in the Friedmann
cosmology with $k=-1$,
we assume that,
in the matter-dominated era,
$a(t)$ satisfies the following differential
equation as in the standard Friedmann cosmology
with $k=-1$ :
\beq
2\, \frac {\ddot{a}(t)}{a(t)} +
\left( \frac{\dot{a}(t)}{a(t)}\right)^2
-\frac{1}{a^2(t)} \simeq 0 ~~.
\eeq
Here, only one of the two initial conditions
can be fixed as $a(t=0)=0$.
The solution of Eq.(38) may be parameterized by an angle, $\Psi$,
as
\beqarr
a(\Psi)	&=&	\alpha t_0 [\cosh \Psi -1]
\\ \no
t(\Psi)	&=&	\alpha t_0 [\sinh \Psi -\Psi]
{}~,
\eeqarr
where $t_0$ is the age of the  Universe and $\alpha$
is a dimensionless parameter to be determined.
In the following, $\Psi_0$ is defined as $t(\Psi_0)=t_0$,
which would correspond to $(1-q_0)/q_0$  in the Friedmann cosmology
with $k=-1$, where $q_0$ is the deceleration parameter.
For illustrative purposes,
we consider, in this Section,
a simple case where the arbitrary function $b(t)$ behaves as
\beq
b(\Psi)	=	\beta \Psi~~,
\eeq
where $\beta$ is a dimensionless parameter to be treated
as a perturbation.
Then, from Eqs.(35) and (37), we have
\beq
 \frac{a(t)}{a(t_0)} = \frac{1}{1+z} \left[
	1+\frac{\beta}{2} \{ \sinh(\Psi_0-\Psi)-(\Psi_0-\Psi)\}
		\right]
	+ {\cal O\/}(\beta^2)
\eeq
and
\beqarr
\Phi(\Psi,\Psi_0)	&=&
\frac{1}{2} (\Psi_0-\Psi)
\\ \no
& & -\frac{\beta}{4} \left[ \cosh
(\Psi_0-\Psi)
+\Psi\sinh (\Psi_0-\Psi)-\frac{\Psi_0^2-\Psi^2}{2}-1\right]
	+ {\cal O\/}(\beta^2)~~.
\eeqarr

Now, we are ready to obtain the redshift--luminosity relation
based on
$d_L=r R(t_0,0)(1+z)$.
Recalling $r(t,t_0)=2\tanh\Phi(t,t_0)$,
$(\Psi_0-\Psi)$ should be expressed in terms of the redshift.
{}From the parameterized solution Eq.(39), $\Psi$ is
\beq
\Psi	=	\cosh^{-1}
		\left[ 1+(
		\cosh \Psi_0-1
		)\frac{a(t)}{a(t_0)}
			\right]~~.
\eeq
Substituting Eq.(41) into Eq.(43), we find
\beqarr
\Psi_0-\Psi
&=&	\Psi_0-\cosh^{-1} \left[1+\frac{ \cosh \Psi_0-1}{1+z}
		\right]
\\ \no
& &
	-\frac{\beta}{2}
		\sqrt{\frac{ \cosh \Psi_0-1} {1+\cosh\Psi_0 + 2 z} }
	\left\{ \sinh (\Psi_0-\Psi)- (\Psi_0-\Psi)
		\right\}
	+ {\cal O\/}(\beta^2)~~.
\eeqarr
For simplicity, we define the zeroth order term of $\Psi$,
$\Psi^{(0)}$, as
\beq
\Psi^{(0)}	=
	\cosh ^{-1} \left[ 1+\frac{ \cosh{\Psi_0}-1}{1+z}
			\right]~~.
\eeq
Then, from Eqs.(42), (44) and (45), we have
\beqarr
\Phi(\Psi,\Psi_0) &=&
	\frac{1}{2} (\Psi_0 - \Psi^{(0)})
\\ \no
& &	+ \frac{\beta}{4}
	\sqrt{\frac{ \cosh \Psi_0-1} {1+\cosh\Psi_0 + 2 z} }
	\left[ (\Psi_0 - \Psi^{(0)})
		-	\sinh (\Psi_0 - \Psi^{(0)})
				\right]
\\ \no
& & 	- \frac{\beta}{4} \left[
	\cosh (\Psi_0 - \Psi^{(0)})
		+\Psi^{(0)}\sinh (\Psi_0 - \Psi^{(0)})
		-\frac{ \Psi_0^2-(\Psi^{(0)})^2}{2}-1
			\right]
\\ \no
& & +{\cal O\/} (\beta^2)
\\ \no
&\equiv&	\Phi^{(0)}(z)+\frac{\beta}{4} \Phi^{(1)}
+ {\cal O\/} (\beta^2)~~.
\eeqarr
In our neighborhood ($z\ll 1$),
$\Phi^{(0)}(z)$ becomes
\beq
\Phi^{(0)}	=	\frac{1}{2}
		\sqrt{ \frac{\cosh \Psi_0-1}{\cosh \Psi_0 +1}}
		\left[ z -
			\frac{ \cosh \Psi_0+2}{2(\cosh\Psi_0+1)}
			\,z^2 \right]
+	{\cal O\/}(z^3)~~.
\eeq
Using Eqs.(44), (45) and (46),
it can easily be seen that
$\Phi^{(1)}$ is zero up to the second order of $z$.
Finally, $r(t,t_0)$ is expressed in terms of the redshift by
\beqarr
r(t,t_0)	&=& 2 \tanh \Phi(t,t_0)
\\ \no
& = &	2 \tanh \Phi^{(0)}(z)
	+\frac{\beta}{2}\Phi^{(1)}(z)
		(1-\tanh^2 \Phi^{(0)}(z)) +{\cal O\/}(\beta^2)~~
\\ \no
&=&	\sqrt{ \frac{ \cosh \Psi_0 -1}{\cosh \Psi_0+1} }
	\left[ z-\frac{ \cosh \Psi_0+2}{2(\cosh \Psi_0+1)}
	\, z^2 \right] + {\cal O\/}(z^3, \beta^2)~~.
\eeqarr
If $\cosh \Psi_0=(1-q_0)/q_0$ as in the Friedmann cosmology,
the standard result,
$r = \sqrt{1-2q_0}[z-(1+q_0)z^2/2]+{\cal O\/}(z^3)$,
is recovered,
implying that the luminosity distance, $r R(t_0)(1+z)$,
has the same dependence on $z$ for the small redshift
as in the standard model.

To summarize, the standard results of the redshift--luminosity
distance relation in the Friedmann cosmology for $z \ll 1$
are preserved in this model,
regardless of whether the local Universe is assumed to
be open or flat.

\section{Numerical Results}

As was demonstrated in the previous section,
the most fundamental cosmological relation predicted by
this model, i.e.,
the redshift--luminosity relation,
is practically the same as in the standard cosmology
for small $z$. This is due to our plausible assumption that $a(t)$
behaves more or less the same as the scale factor in the standard
cosmology.
We now proceed to explore the relationship for large $z$.
Unfortunately,
it is impossible to do so analytically,
due partly to our
inability to analytically solve the light propagation equation.
In this Section, therefore, we resort to
numerical method and present some numerical answers in the form
of figures depicting
possible modifications
of the redshift--luminosity distance relation for large $z$.
Since  our local Universe appears to be open as discussed before,
we only consider the second case of the previous section.
Thus, $a(t)$ is parameterized by an angle $\Psi$ as given in Eq.(39).
Since $b(t)$ is completely arbitrary except for being small,
we consider the following three representative
cases : (A) $b(\Psi)=\beta \Psi$,
(B) $b(\Psi)=\beta \Psi^4$ and
(C) $b(\Psi)=\beta [\cosh \Psi - 1]^2$,
where $\beta$ is a dimensionless parameter to be determined.
Note that the case (C) corresponds to the picture in which
there is a $constant$ vacuum energy density in the  Universe,
as can be seen from Eq.(17).
Recalling that the Hubble expansion rate of the local Universe
at the present epoch, $\overline{H}_0\equiv \dot{R}/R|_{t=t_0,r\sim0}$,
is $[ \dot{a}(t)/a(t)]_{t=t_0}$,
the density parameter in our neighborhood is, from Eq.(18),
\beq
\overline{\Omega}_0
	\equiv \frac{8 \pi  G \rho(t=t_0,r\simeq 0)}
		{3 \overline{H}_0^2}
	=	1- \dot{a}_0^2 - b(t_0)\dot{a}_0^2~~,
\eeq
where the bar denotes the $local$ value and the subscript
zero represents the $present$ value.
With the definitions of
$ \overline{\Omega}_{0,a} \equiv 1-\dot{a}_0^2$ and
$\overline{\Omega}_{0,b} \equiv b_0 \dot{a}_0^2 $,
the density parameter and the pressure of the local Universe
can be expressed by
\beqarr
\overline{\Omega}_{0}
&=& \overline{\Omega}_{0,a}
-\overline{\Omega}_{0,b}~~,
\\
8 \pi\overline{ Gp_0} &=&\overline{\Omega}_{0,b}
\overline{H}_{0}^2~~,
\eeqarr
where Eq.(19) is used.

{}From the observation of small peculiar velocities of nearby galaxies,
we assume that  the pressure of the local Universe is
relatively small compared with the energy density.
That is, $\overline{\Omega}_{0,b} $ is very small.
In the numerical calculations to be presented in this Section,
we assume $\overline{\Omega}_{0} =0.1$  and, for the sake of
definiteness, choose values
$\overline{\Omega}_{0,a} =0.1001$ and
$\overline{\Omega}_{0,b} =0.0001$, specifying the numerical values
of $\alpha$ and $\beta$.
The light propagation equation can then be numerically
integrated with the boundary condition $r(t=t_0, t_0)=0$.
To this end, we divide the numerical solution of
$r(t,t_0)$  into $N$ intervals and fit each interval using
\beq
r_i(t,t_0)= \delta_i \left[ t_0^{\gamma_i}
-t^{\gamma_i} \right]~~\mbox{ for\/}~~
\frac{i-1}{N} < \frac{t}{t_0} < \frac{i}{N}~~
(i=1,2,3...)~,
\eeq
which then yields the numerical values of $\delta_i$ and $\gamma_i$
for each interval. We have used $ N = 200$ in our numerical calculations.
Using the definition of the comoving distance, i.e.,
$r_i(t,t_0)=r_i(t+\Dt,t_0+\Dtz)$, we have the redshift,
for each interval, as
\beq
(1+z) \equiv \left( \frac{\Dtz}{\Dt} \right)_i
	=	\left( \frac{t}{t_0} \right)^{\gamma_i-1}
{}~.
\eeq
Then, the luminosity distance, $d_L=r R(t_0)(1+z)$,
for each interval, can easily be calculated from Eqs.(52) and (53).
In Fig.1, the results of the three cases discussed above
are presented by solid lines,
while the standard results with $\Omega=0.1$ and $\Omega=1.0$
by the dashed lines and dotted lines, respectively.
First, it is to be noted that on small scale, the
linear {\em Hubble diagrams\/} are preserved for all of the
three cases, as was shown by perturbative calculations.
In the cases of (A) and (C),
the redshift--luminosity distance relations in this model
are almost indistinguishable
and furthermore correspond to
 the standard results with the density parameter
in the range $\Omega \sim$ $0.8-0.9$,
even though the local mass density  in the model is set
to be $\overline{\Omega}_0=0.1$.
In the case (B), however, it appears to be the standard result
with $\Omega > 1$ although the difference is not very
significant. Nevertheless,
this qualitative feature was totally unexpected.
In this model, therefore, a
precise measurement of the redshift--luminosity
distance relation alone cannot provide information on  $q_0$,
which is related to $\Omega_0$ as $2q_0=\Omega_0$,
contrary to the case of the standard cosmology.

Now it would be appropriate to discuss the meaning of the
$observed$ increase of $\Omega_0$ as we look farther out.
Every light signal we are receiving right now
contains information about the past in time.
That is, what we measure are,
for example, the redshift and $G\rho(t)$, not $G\rho(t_0)$.
Thus, we deduce physical quantities at the present time $t_0$,
using the standard cosmological evolution equations.
Recalling that $G\rho(t)$ in the matter dominated era in the standard
Friedmann cosmology is proportional to $1/S^3(t)$,
and $S(t)$ is just $S(t_0)/(1+z)$,
$G\rho(t_0)$ is $obtained$ from $G\rho(t)$ by multiplying $1/(1+z)^3$.
Dividing $G\rho(t_0)$ with the constant critical density
in the standard cosmology, $G\rho^{(s)}_c$,
we can deduce the density parameter at present time,
which we shall call $\Omega^{obs}_0$, as
\beq
\Omega^{obs}_0 \equiv  \frac{G\rho^{obs}(t_0)}{G\rho^{(s)}_c}
	\equiv \frac{G\rho(t,r(t))}{G\rho^{(s)}_c (1+z)^3}~~.
\eeq
Using Eqs.(52) and (53) for each interval in Eq.(18),
we have calculated $\Omega_0^{obs}$ versus the redshift.
The results are shown in Fig.2. As before, we have taken
the local value $\Omega_{0}^{obs} =0.1$. In all three cases,
the calculated values of $\Omega^{obs}_0$ are increasing
 functions of $z$. ( A naive value in the standard cosmology
is supposed to be a constant.)
We can see that the cases of (A) and (C) are
practically identical, whereas
 the case (B) shows a faster increase of $\Omega_0^{obs}$.
As can be seen in Fig.2, however, even in the case (B) the increase
is too slow to explain  the IRAS data\cite{IRAS}
where $\Omega^{obs}_0$ is compatible to unity
(but with large errors) at the distance of about
several 100 Mpc($z \sim 0.0166$).
To fit the increase of $\Omega^{obs}_0$ up to unity at $\sim 100$ Mpc
 requires  drastic ( perhaps unrealistic ) changes
in the form  of and parameters in $a(t)$ and $b(t)$,
which then would modify our predictions on the Hubble law.

\section{summary and conclusions}

We have studied how  cosmological observables are modified
in an isotropic but inhomogeneous Universe
compared with those of the standard model.
In particular, the luminosity distance and the density
parameter  as functions of the redshift have been examined
in the generalized Robertson--Walker spacetime with only one
$(t,r)$-dependent scale factor and
they were compared with the standard results.

When $R(t,r)$ is not  factorized  into the form of
$a(t)f(r)$, the simple
redshift--scale factor relation such as $(1+z)=a(t_0)/a(t)$
remains no longer valid.
First by solving  light propagation equation, Eq.(2),
for radially propagating light with
the  boundary condition $r(t=t_{received})=0$
and then considering two wave crests emitted at time $t$ and $t+\Dt$
which are received at $t_0$ and $t_0+\Dtz$, respectively,
we have obtained the general redshift--scale factor relation given by
Eq.(5).
The result is valid in an inhomogeneous Universe and is shown to be
reduced to the simple form $(1+z)=a(t_0)/a(t)$ in the case of
the homogeneous spacetime, i.e.,  the standard
Robertson--Walker spacetime.
Our general relations agree with the
results obtained in \cite{Bondi} and \cite{Moffat}.

We have applied the general redshift--scale factor relation
to the cosmological model in \cite{Kim}
where the Universe is pictured as being
 inside a highly dense and rapidly expanding shell
with the underdense center.
First, for the nearby objects ($z\ll 1$),
the luminosity distances  as  functions of the redshift
are obtained analytically,  using a perturbative method
for two cases where the underdense center is either flat or open
according to the definition of the standard Friedmann cosmology.
One of the most interesting features in \cite{Kim}
is that the scale factor $R(t,r)(=a(t)/(1-B(t)r^2) )$
is specified by two arbitrary functions, $a(t)$ and $B(t)$
(or $b(t)$), and $a(t)$ is very similar to the scale factor
of the standard Friedmann cosmology and $B(t)$ (or $b(t)$)
is the perturbation to the locally flat (or open) Universe.
Under the assumption that $a(t)$ behaves the same as that in the
standard cosmology,
it is shown analytically  that the standard redshift--luminosity
distance relations in the Friedmann cosmology for $z \ll 1$ remains intact
for both cases.
Specifically,
since the corrections of order $\cal{O \/}(\beta)$
of these expressions can
be expanded as a power series of $\Sigma_{i=3}^{\infty} c_i z^i$
with some coefficients $c_i$ (that is,
zero up to the second order of $z$),
it has been shown that for nearby objects, in spite of its
different metric,
the cosmological model of \cite{Kim}
is not much different from the standard cosmology as far as
the Hubble law is concerned.
It is also interesting to note that
in spite of the special location of the observer, i.e., the return of
the pre--Copernican notion in the model \cite{Kim},
the results are almost the same as those of the standard model.

As for large $z$, the redshift--luminosity distance relations
given in Eqs. (32) and (48) are different
from   those of the standard
cosmology and moreover it is expected
that the corrections would be larger when
${\cal O\/}(\beta^2)$ terms are included. In this case, as we
mentioned repeatedly, we cannot
use the perturbative method and thus
we have solved them numerically and obtained the results as shown
in Fig.1. and Fig.2.

Figure 1 shows the redshift--luminosity distance relation
in the cosmological model of \cite{Kim}.
Comparing them with the standard cosmology with $k=0$ (dotted curve)
and $k=-1$ (dashed curve),
we can easily see that for small $z$,
the redshift--luminosity distance relation of model \cite{Kim}
denoted by the solid lines
is  almost the same as the standard one,
as was also  shown in the explicit perturbative calculation.
But for $1<z<3$, the Hubble law of the model is very
similar to  that of the standard cosmology with $k=0$,
not with that with $k=-1$,
in spite of the fact that the mass density of the
local Universe is set to be
$\overline{\Omega}_0=0.1$.
It is to be noted  that although there is no substantial
deviation from the standard model
in our redshift--luminosity distance relation in Fig.1,
the deviation in the case of the TB model is more
prominent and is different from ours, both qualitatively and
quantitatively  as was shown in \cite{Moffat}.
Figure 2 shows that the calculated density parameter
 at the present time,
$\Omega^{obs}_0$, is an increasing function of the redshift.

Although, admittedly,  our perturbative calculations may not be
rigorous in the sense that the functions $a(t)$
and $B(t)$ that appear in the scale factor
$R(t,r)$ could not be determined by
the  equation of state, and furthermore
our numerical results can only explain  the IRAS data qualitatively,
we feel that the qualitative
nature of our results are robust
because we have considered the general case with
$B(t)=\beta (t/t_0)^n ~(n >0)$,
$b(t)=\beta \Psi(t)$, $b(t)=\beta \Psi^4(t)$ and
$b(t)=\beta [ \cosh \Psi -1]^2$
with a variety of numerical values of the parameters involved,
all of which have led to similar results.

In summary, although the scale-dependent cosmology for
the inhomogeneous Universe as modeled in
\cite{Kim} implies the explicit running of $H_0$, $\Omega_0$ and $t_0$
as functions of $r$ because of the non-Robertson-Walker metric,
as far as the observables such as redshift--luminosity
distance relations are concerned, the results are
hardly  different from those of the
standard model  in our neighborhood, i.e. for small $z$.
Even for large $z$, the difference between the model
considered and the standard model with $k=0$
still remains small  but
the model can be tested  when the data from
galaxy redshift survey at long distance become available
and are compared with the predictions of the model on the
matter distribution and on the age of the Universe.\cite{Kim}

\acknowledgments
The authors would like to thank G. Feldman, M. Im
and Y.C. Pei for helpful discussions.
One of the authors (THL) wishes to thank
the Department of Physics and Astronomy,
the Johns Hopkins University for the hospitality extended
to him during the
completion of this work.
This work was supported in part by the Basic
Science Research Institute Program,
Ministry of Education, Korea,
Project  No. BSRI-94-2418(THL) and by the National Science Foundation.

\newpage
\begin{center}
\begin{large}
\begin{bf}
Figure Captions
\end{bf}
\end{large}
\vspace{5mm}\
\end{center}
\begin{description}
\item [Fig. 1]
The luminosity distance, $d_L$,
as a function of the redshift $z$ for the  three cases
 (A), (B) and (C) as discussed in the text.
Solid lines denote the results in our model,
whereas dashed and dotted lines  indicate the standard results
with $\Omega=0.1$ and $\Omega=1.0$, respectively.

\item [Fig. 2]
The calculated density distribution, $\Omega^{obs}_0(z)$
as a function of the redshift $z$.
The numerical results for the cases (A) and (C)
are practically indistinguishable.

\end{description}
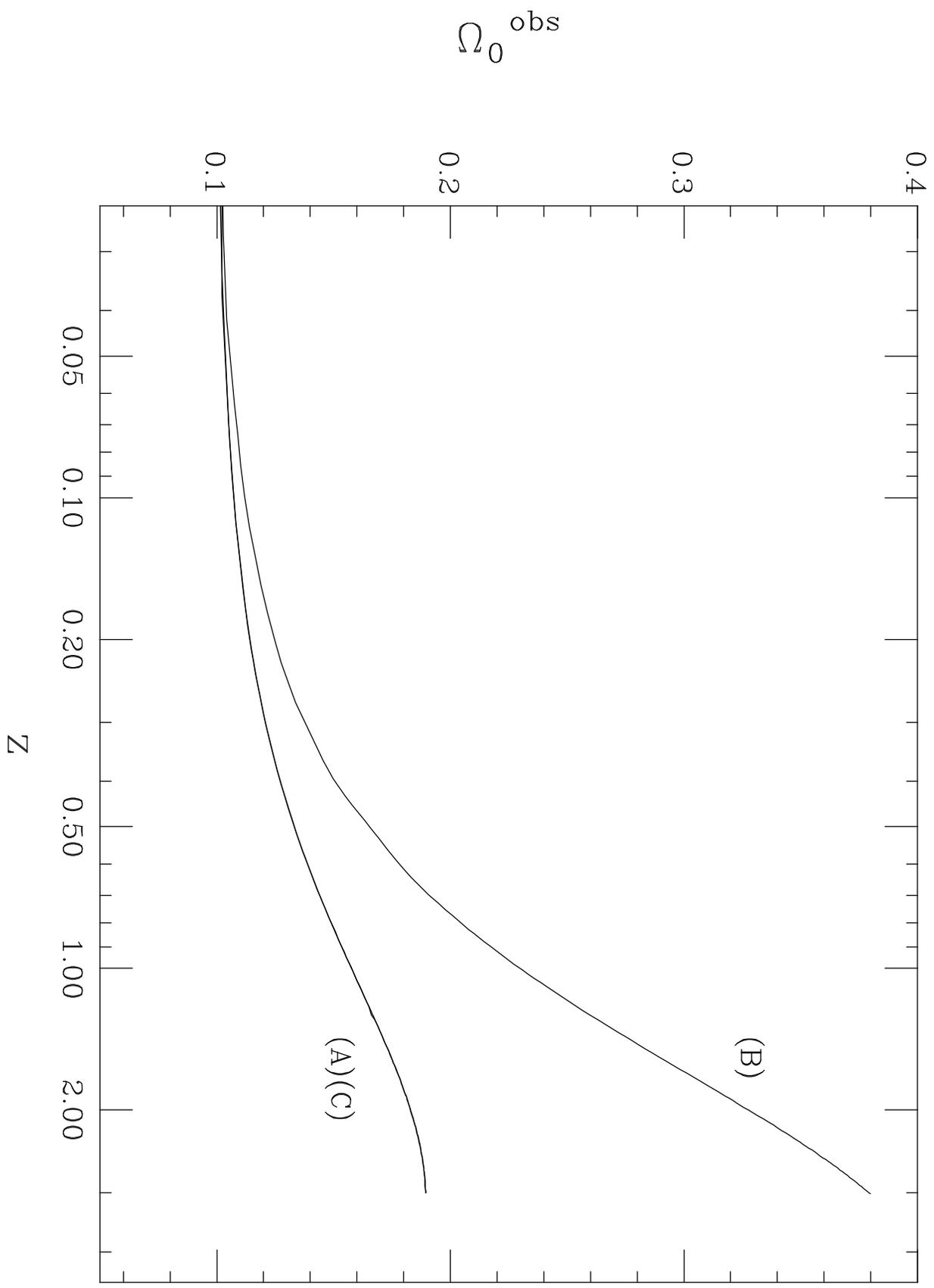
\end{document}